Title: The London model of Van-der-Waals forces: what's next?
Author: Mladen Georgiev {Institute of Solid State Physics, Bulgarian Academy of Sciences, 1784 Sofia, Bulgaria}
Comments: 9 pages incorporating 3 figures, all pdf format
Subj-class: physics

London's polarization model is extended over a wide range of VdW attraction. At low temperature the VdW attraction is the main competitor to the Casimir force as there seems to be more than one exchange agent. As the temperature is raised, the VdW force is gradually decreasing by virtue of a gradually falling down polarizability. We also show that the VdW coupling is the chief component of cohesion by an universal force at the initial stages of the buildup of matter. The electronic polarizability, accounting for the conventional approach to polarization phenomena, result in polarizabilities of the order of e.g. a lattice constant This trend is advanced in multilevel systems, such as the off-center defects, where the polarizability is renormalized through electron-vibrational mode coupling to gain a multiple vibronic enhancement exceeding by many orders of magnitude the bare electronic estimate. In concomitance, we distinguish between centrosymmetric species, such as the s-atoms, and non-centrosymmetric species of lower symmetry. Applications to high-temperature superconductivity and cosmology are discussed.

1. Preface

The Van der Waals (VdW) force between atoms and molecules have long occupied the attention of both theorists and experimentalists[1], [2], [3]. The simple though transparent physical model elaborated by Fritz London for easy conclusions [4] has helped to establish the basis for more sophisticated discussions. Recently, the discussion has been revived in view of the nature of low-temperature force Casimir or VdW [5]. In what follows, we describe the fundamentals of the model in the order of increased complexity. Priority is given to novel physical hypotheses which are introduced only after a due phys and math consideration.

2. F. London's model

Herein we follow London's considerations as reported in terms of a quasiclassical language [4]. Following this we later pay due attention to a more vigorous quantum mechanical approach. The origin of the VdW dispersion forces is obliged to the occurrence of significant instantaneous dipole moments in atoms where the average dipole moments are vanishing (s-atoms). These instantaneous dipoles couple through dispersive coupling on a time scale that is premature to the occurrence of any longer term interactions. The dispersive coupling between neighboring atoms occurs to synchronize the orbital motion of electrons in nearby atoms and thereby the occurrence of coupling.

By a simple line of reasoning, we consider two s-atoms whose positive nuclei are firmly fixed, while their respective outer electrons vibrate along X about their equilibrium positions. The dipole associated with the electronic oscillator (quasielastic oscillator) is $\mu = ex = \alpha E$ where $x$ is the displacement, $\alpha$- the dipolar polarizability, $E$- the electric field. It follows that for a quasielastic oscillator $eE = Kx$ and the quasielastic force constant is $K = eE/x = e^2/\alpha$, m is the electron mass. It follows that the vibrational frequency of a quasielastic oscillator is $\omega = \sqrt{(e^2/\alpha m)}$.

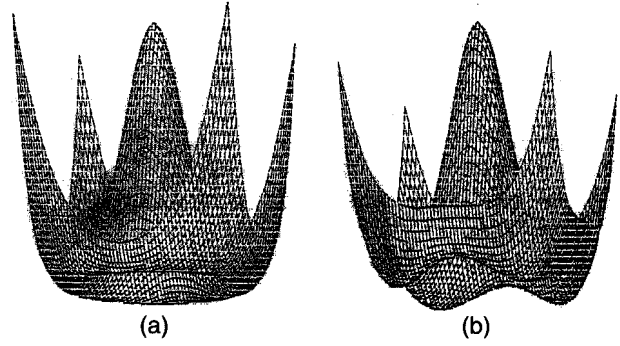

(a) (b)

Figure 1: Adiabatic sombrero-type potentials by linear electron-phonon coupling (a) and up to cubic electron-phonon coupling (b). The barriers appearing on the lower surface in (b) are clearly distinguishable.

Following easy manipulations and reasoning, the complete energy of a system of two oscillators is found to be

$$\varepsilon = (1/2m)(p_1^2 + p_2^2) + (e^2/2\alpha)(x_1^2 + x_2^2) - (2e^2 x_1 x_2 R^3) \qquad (1)$$

R is the separation of the two oscillators along their interconnecting axis. Introducing the symmetric and antisymmetric coordinates, the tunneling splitting of the oscillators is

$$\Delta\varepsilon = (1/2m)p_s^2 + [(e^2/2\alpha) + (e^2/R^3)]x_s^2 + (1/2m)p_a^2 + [(e^2/2\alpha) - (e^2/R^3)] x_a^2 \qquad (2)$$

with split off symmetric and antisymmetric frequencies

$$\omega_s = \sqrt{\{(e^2/m\alpha)[1 + (2\alpha/R^3)]\}}, \quad \omega_a = \sqrt{\{(e^2/m\alpha)[1 - (2\alpha/R^3)]\}} \qquad (3)$$

The total zero-point oscillator energy is found to be

$$\varepsilon_0 = \tfrac{1}{2} h (\nu_s + \nu_a) = h\omega [1 - (\alpha^2/2R^6)] \qquad (4)$$

all at $R \gg |x_1|, |x_2|$ .

3. VdW interactions by 2$^{nd}$ order perturbation polarizabilities

3.1. Bandgaps in phonon-coupled- or in decoupled- two-level centrosymmetric systems

For a system in equilibrium with an external field **F**, the expansion of the total energy in the field variables is of importance. We use the field expansion:

Figure 2: Calculated examples of azymuthal φ-dependences of Mathieu's periodic eigenfunctions, odd parity $se_m(z,q)$ and even parity $ce_m(z,q)$ using the series expansions at "not too large q", as given in Appendix. The abscissa is $Z = (2/\pi)z = (4/\pi)\varphi$.

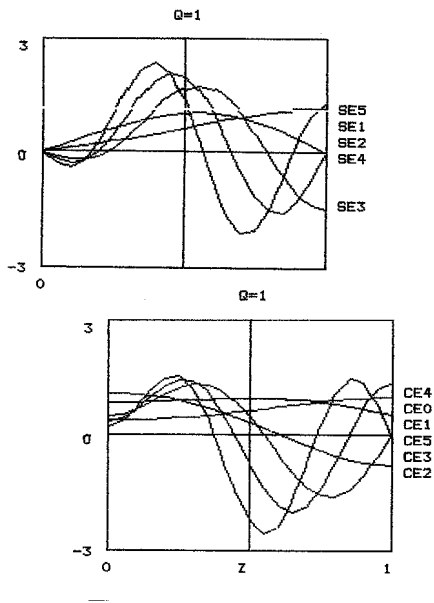

$$E(\mathbf{F}) = \Sigma_{ij}\alpha_{ij}F_iF_j$$

where $\alpha_{ij}$ is the polarizability tensor. It should be noted that the first derivative of $E(\mathbf{F})$ is vanishing because of the field→system equilibrium condition. Under it only 2$^{nd}$ order perturbation theory is meaningful for deriving the polarizability $\alpha_{ij}$. It gives

$$\alpha_{ij} = \Sigma_m \langle n| p |m\rangle\langle m| p |n\rangle / (E_n - E_m)$$

$$= p_{mn}^2 / (E_n - E_m) = p_{mn}^2 / \Delta_{mn} \qquad (6)$$

for a 2-level uncoupled system with levels m and n and transition dipole $p_{nm}$.

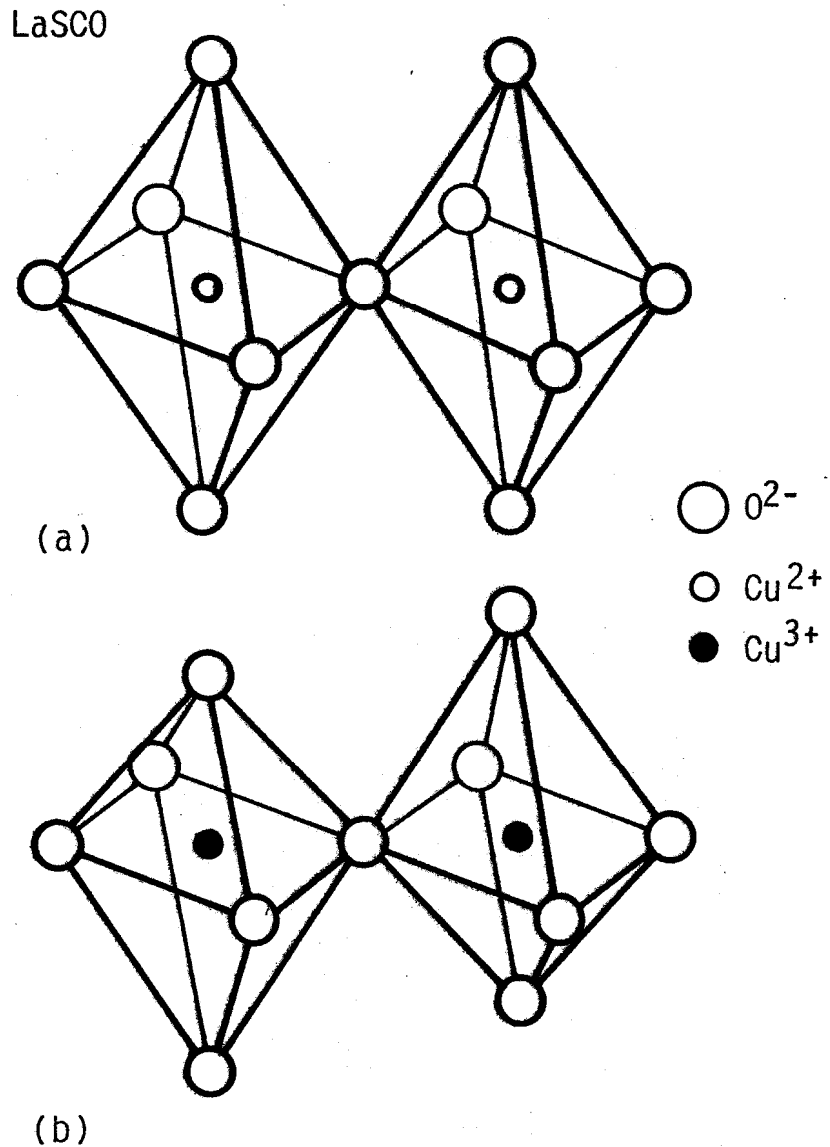

Figure 3: Pairing of nearest-neighboring Cu-O octahedra by inversion-dipole coupling in $La_{2-x}Sr_xCuO_4$ superconductor, decoupled in (a) and coupled in (b).

Here the quantity $\Delta_{mn}$ is the energy bandwidth of Holstein's polaron [6]. $\Delta_{mn}$ renormalizes to $\Delta_{mn}^{ren} = \Delta_{mn} \exp(-E_{JT} / 2\hbar\omega_{ren})$ where $\omega_{ren}$ is the renormalized mode frequency at the 2-level system, $E_{JT}$ is the Jahn-Teller energy. The renormalized exponential $\exp(-E_{JT} / 2\hbar\omega_{ren})$ is known as Holstein's narrowing factor in so far as it brings in a narrowing of the original level gap. At standard values of the entering parameters the narrowing could amount to orders of magnitude. If so, Holstein's narrowing may lead to substantial squeeze of the 2-level bandgap and, therefore, to orders of magnitude elevations of the 2nd order-renormalized (vibronic) polarizability exceeding largely the bare electronic value. In as much as renormalization requires small perturbations, the problem should be readdressed in order to separate real physics from math artifacts.

3.2.. VdW force in coupled or decoupled systems with 2$^{nd}$ order perturbation polarizability

Following closely London's model, we set in general terms the VdW force arising in a two-level system by the interaction of fluctuation dipoles arising from the polarizability:

$$\Phi(R)^{decoupled} = \Delta_{mn} (\alpha_{mn} / \kappa R^3)^2 = (p_{mn}^4 / \kappa^2 R^6) / \Delta_{mn} \qquad (7)$$

$$\Phi(R)^{coupled} = \Delta_{mn}^{ren} (\alpha_{mn} / \kappa R^3)^2 = (p_{mn}^4 / \kappa^2 R^6) / \Delta_{mn}^{ren} \qquad (8')$$

$$\Phi(R)^{colossal} = [(p_{mn}^4 / \kappa^2 R^6) / \Delta_{mn}] \exp+(E_{JT} / 2h\omega_{ren}) \qquad (8'')$$

Here $\Delta_{mn} / h$ is the average frequency of circumventing the barrier interposing two neighboring equilibrium positions of the double-well system, R is the intermolecular separation. The latter line referring to a coupled system contains the reciprocal of Holstein's narrowing factor. With a realistic Jahn-Teller energy of 25 meV and phonon frequency of 2.5 meV we get for the narrowing factor $\exp+(E_{JT} / 2h\omega_{ren}) \sim \exp(10) \sim 84$. The resulting colossal force has been predicted on grounds of the 2$^{nd}$ perturbation expression and may not at all hold true for large forces lifting the ($|\Phi| \ll 1$) limitations of perturbation theory.

We should note at this point that lifting the 2$^{nd}$ order limitations, not as easy though, may open the way of implicating the theory to wider range interactions such as the quadruple coupling. Generally, the multipole expansions have been discussed at length elsewhere [7].

3.3. Inversion electric dipoles in off-centered non-centrosymmetric systems

The off-center species violate the inversion symmetry of a system. They enter into equation (8) to give rise to site-splitting and to inversion-dipole VdW force. Both should be described separately, though by similar mechanisms. The Hamiltonian of an off-center system has been discussed elsewhere [8]. The phonon coupling comes as phonon mixing of two opposite-parity electronic states by an odd-parity vibrational mode. In concomitance, the interlevel energy gap $\Delta_{mn}$ is now composed of opposite parity components and so is its squeezed polaron narrowed bandgap $\Delta_{mn}^{ren}$.

The eigenstates have not been derived in general though they certainly have in 2D. The 2D eigenfunctions are Mathieu's periodic functions [9]. The quantum mechanical pattern attached to them spread (smeared) all around the central peak. The system is polarizable in a fluctuating dipolar field as before and gives rise to an inversion dipole and a dispersive force to couple to any nearby off-center entity.

Although the central VdW force has been studied extensively [3], little has been said about its off-central inversion alternative. It is therefore feasible to comment on both of them in order to compare and assess the prospects for materializing any under reasonable conditions. It is beyond any doubt that the renormalized bandgap $\Delta_{mn}^{ren}$ can be composed in a way similar to

Holstein's polaron band by factorizing $\Delta_{mn}^{ren}$ to Holstein's exponential narrowing factor. Only, in lieu of Holstein's polaron bands, original and squeezed, we should now have polaron bandgaps, original and squeezed.

4. 2D dispersive interactions by Mathieu functions in non-centrosymmetric systems

In looking for better predictions of London's model to cover a wider range of the energy scale, we apply his approach to the polarizability and the VdW force though expressed by the exact wavefunctions to calculate 2D dipoles and eigenenergies. This is expected to bring in improvements to the symbolistics, though still retaining the typical uncertainties characteristic of the 2$^{nd}$ order approach. Most of all this refers to equations (6) through (8), where we do not really know how to express the polarizability and the dispersive interaction. Therefore, we use the 2$^{nd}$ perturbation expressions as a formal basis to adding corrections accounting for a more refined approach. Figure 2 shows a double-well sombrero potential in 2D.

Because of tunneling across the symmetry barrier, non-centrosymmetric systems regain its former symmetry on the average though with a larger radius and reactivity. So, what is lost by mixing is regained by tunneling. This is a profound feature of the non-centrosymmetric system due to smearing. The smeared system is generally larger and easier to comprehend. Figure 3 shows selected periodic Mathieu functions.

Due to ion smearing around the central barrier site, the effective orbital by Mathieu's function increases its radius several times which leads to a large quadratic increase of the active area under the off-center volume. As a result of the volume increase, the off-center ion reactivity is largely increased and so is its dispersive coupling with the off-center ions nearby. All this makes the off-center system largely competitive to on-center alternatives in the quest for higher polarizabilities and stronger dispersive coupling.

5. Temperature dependence of non-centrosymmetric polarizability

Assuming Maxwell-Boltzmann statistics for the polarizable two-level species in an external electric field:

$n_{\pm}(k_BT) = n_0 \exp(\pm e\varphi / k_BT)$

we get for the 1D polarizability (coupling to the second power of field) [10] $\alpha(k_BT) = \alpha_0$ [exp+($E / k_BT$) − exp-($E / k_BT$)] / [exp+($E / k_BT$) + exp-($E / k_BT$)]

$= \alpha_0 \tanh[(\Delta / k_BT)]$ (9)

where $\Delta$ is the tunneling splitting. At low temperatures ($e\varphi \ll k_BT$), $\alpha_0 = \alpha(0) = p_0^2 / 3\Delta$. The case of Fermi-Dirac statistics will be dealt with separately. We note that the quantum F-D statistics turns nearly classic in the Boltzmann-Tail range.

Another important theoretical result is the 2D- polarizability which comes closer to important experimental situations with essential implications for the planar species [10 Bersuker]:

$\alpha(k_BT) = (p_0^2 / 3k_BT) [ \exp(- \Delta / k_BT) + (k_BT / \Delta) \sinh(\Delta / k_BT) ] /$

$$[ \exp( - \Delta / k_BT) + \cosh( \Delta / k_BT) ] \tag{10}$$

where $\Delta$ is the 2D tunneling splitting and $p_0$ is the mixing dipole as above (four adiabatic potential minima along off-center circle).

For fermionic systems, the approach based on Fermi-Dirac's statistics is a better guess. We introduce the fermion distribution function

$$f(E_n) = 1 / \{ [1 + \exp[(E_n - E_F) / k_BT] \} \tag{11}$$

where $E_n$ is the quantized energy level, $E_F$ is Fermi's energy (chemical potential). The Boltzmann tail obtains at $E_n > E_F$ for $k_BT > (E_n - E_F)$ which gives $f(E_n) \sim \exp[-(E_n - E_F) / k_BT]$. At the other end,. $E_F > E_n$ and $k_BT < |(E_n - E_F)|$ the distribution function $f(E_n) \sim 1$.

6. VdW pairing of non-centrosymmetric characters

A simple BCS formula for calculating the critical temperature of superconductivity [11]

$$k_BT_c = 1.14\, \hbar\omega_D \exp( - 2 / \lambda ) \quad \text{with} \quad \lambda = N(0)V, \tag{12}$$

where V is the pairing potential, N is the density of states, and $\omega_D$ is Debye's frequency, applies to pairing in momentum space at weak coupling leading to Cooper pairs. Applying the colossal VdW modification results in real space pairing through VdW bipolarons and bipolaronic superconductivity through equations (7) through (8), respectively:

$$V(R)^{decoupled} = (p_0^4 / \kappa^2 R^6) / \Delta_{mn} \quad \text{(BCS momentum-space pairing)}$$

$$V(R)^{coupled} = [(p_0^4 / \kappa^2 R^6) / \Delta_{mn}] \exp+(E_{JT} / 2\hbar\omega_{ren}) \quad \text{(Bipolaronic real-space pairing)}$$

Under all other conditions identical, the ratio between coupled and decoupled pairing energies amounts to $\exp+(E_{JT} / 2\hbar\omega_{ren})$ which may imply some 84 fold increase of pairing power. The Coulomb repulsion is irrelevant in polarization phenomena.

7. Nature of the low-temperature force: VdW or Casimir?

In scientific literature, it is customary to raise the question as to the nature of the low-temperature force. Two competitors are mentioned, the quantum mechanical Casimir force and the VdW force in its three forms: decoupled (7), coupled (8') and colossal (8"). For this purpose we compare the low-temperature behavior of the two under compatible conditions. The Casimir force between two parallel metal plates of area $A$ each reads [5]:

$$\Phi_{casimir} = - (hc\pi^2) A / 240\, d^4 = - (hc\pi^3) a^2 / 240\, d^4 = - (4hc\pi^4) a^2 / 240\, d^4 \tag{13}$$

where $d$ is the plate-to-plate separation, $c$ is the speed of light. We set $A = \pi a^2$ where $a$ is the radius of circular plate to get the 2$^{nd}$ row of equation (13). For a spherical plate we fill in the last row. Now we compare eq. (13) therein with $\Phi(R)$ of equation (7) at $R = a$ to give

$\Phi_{decoupledVdW} / \Phi_{casimir} = [(p_0^4 / \kappa^2 R^6) / \Delta_{mn}] / [(4hc\pi^4) a^2 / 240 d^4] \gg 1$

$R^6 = [(p_0^4 / \kappa^2) / \Delta_{mn}] / [(4hc\pi^4) a^2 / 240 d^4] \ll 1$

It may be seen that at the lowest temperatures R emerges good enough to secure $\Phi_{decoupledVdW} / \Phi_{casimir} \gg 1$.

## 8. Concluding remarks

We outlined the peculiarities of a simple London approach to dispersive interaction utilizing phonon coupled electrostatic polarization in solids. The approach is applicable to either attractive or repulsive coupling though our considerations concerned the former mainly. This was made in view of a planned excursion to the expected interactions of colossal strength far exceeding the usual 2nd order perturbation analysis to the electrostatic polarization.

Our conclusions suggested that the theoretical approach to polarization could be widened by including quadruple interactions to the 1st place.

For the 1st time we described the dispersive interactions in an ensemble of off-centered configurational characters, each one carrying a non-centrosymmetric inversion electric dipole. Because of configurational tunneling they transform into highly polarizable spherical species. The emerging picture complements usual analyses based on centrosymmetric species which originally lack the inversion asymmetry. Our result widened the approach to dispersive forces in its own right. A mixture of combined centrosymmetric and non-centrosymmetris original characters has the potential of improving several times the effectiveness of the dispersion theory through enhancing the overall polarizability of the system. In addition, the transition temperature $T_C$ is also increased by virtue of the increase of the DOS N(0).

In closing, we only add the possibility that a dipolar-mode enhancing mechanism may be responsible for thr colossal character of the dispersive VdW coupling, as explained in Ref. [12] and elsewhere.

Finally, it may be noticed that no retardation effects have been accounted for presently. The omission has been made so as not to complicate considerations at this stage.